\documentclass{article}
\usepackage{icrctc07}

\title{Study of time lags in HETE-2 Gamma-Ray Bursts with redshift: search for astrophysical effects and Quantum Gravity signature}
\shorttitle{Study of time lags in HETE-2 Gamma-Ray Bursts}
\authors{J. Bolmont$^{1,4}$, A. Jacholkowska$^{1}$, J.-L. Atteia$^{2}$, F. Piron$^{1}$ and the HETE-2 collaboration.}
\shortauthors{J. Bolmont et al}
\afiliations{$^1$LPTA, Universit\'e Montpellier 2, CNRS/IN2P3, Montpellier, France\\ $^2$LATT, Universit\'e Paul Sabatier, CNRS/INSU, Toulouse, France\\ $^4$Present address: DESY, D-15738 Zeuthen, Germany}
\email{julien.bolmont@desy.de}

\abstract{The study of time lags between spikes in Gamma-Ray Bursts light curves in different energy bands as a function of redshift may lead to the detection of effects due to Quantum Gravity. We present an analysis of 15 Gamma-Ray Bursts with measured redshift, detected by the HETE-2 mission between 2001 and 2006 in order to measure time lags related to astrophysical effects and search for Quantum Gravity signature in the framework of an extra-dimension string model.
The use of photon-tagged data allows us to consider various energy ranges. Systematic effects due to selection and cuts are evaluated.
No significant Quantum Gravity effect is detected from the study of the maxima of the light curves and a lower limit at 95\% Confidence Level on the Quantum Gravity scale parameter of $3.2\times10^{15}$~GeV is set.
}

\begin{document}
\maketitle

\section{Introduction}

Two approaches are followed trying to elaborate a Quantum Gravity theory, which would make the link between particle physics and cosmology: String Theory and Loop Quantum Gravity (which will not be discussed here, see \cite{gambini}). In String Theory, gravitation is considered as a gauge interaction and Quantum Gravity effects result from graviton-like exchange in a background classical space-time.

In this paper, we follow \cite{ellis1,ellis2} and we consider a model \cite{ellis3} based on String Theory where photons propagate in the vacuum which may exhibit a non-trivial refractive index due to its foamy structure on a characteristic scale approaching the Planck length $l \sim m^{-1}_\mathrm{Planck}$. This implies a light group velocity variation as a function of the energy of the photon $E$: $v(E) = c/n(E)$, where $n(E)$ is the refractive index of the foam. Generally, the Quantum Gravity energy scale $\mathrm{E}_\mathrm{QG}$ is considered to be close to the Planck scale. This allows us to represent the standard photon dispersion relation with $E/\mathrm{E}_\mathrm{QG}$ expansion:
$$c^2p^2 = E^2 (1 + \xi \frac{E}{\mathrm{E}_\mathrm{QG}} + O(\frac{E^2}{\mathrm{E}_\mathrm{QG}^2}))\ \ \mathrm{and}$$
$$v(E) \approx c\left(1-\xi\frac{E}{\mathrm{E}_\mathrm{QG}}\right),$$
where $\xi$ is a model parameter whose value is set to 1 in the following \cite{amelino1}. Then, the time lag between two photons with energy difference $\Delta E$ and emitted at the same time is given by
\begin{equation}
\Delta t = \mathrm{H}_0^{-1} {\frac{\Delta E}{\mathrm{E}_\mathrm{QG}}} \int_{0}^{z} \frac{(1+z)\,dz}{h(z)},
\label{form:dt}
\end{equation}
where 
$$h(z) = \sqrt{\Omega_\Lambda + \Omega_\mathrm{M} (1+z)^3},$$
and with $\Omega_{tot} = \Omega_\Lambda + \Omega_\mathrm{M} = 1$, $\Omega_\Lambda = 0.7$ and H$_{0} = 71\:\mathrm{km}\:\mathrm{s}^{-1}\:\mathrm{Mpc}^{-1}$.

This formalism is different from the one used in \cite{ellis1,ellis2}, with the addition of an extra factor \mbox{$(1 + z)$} in the integral of Eq.~\ref{form:dt}.

Gamma-Ray Bursts (GRBs) are well suited for this kind of study. They are very bright sources at cosmological distances, and their light curves have spikes that can be used to measure time lags. In order to probe the energy dependence of the velocity of light induced by Quantum Gravity, we analyse the time lags of 15 GRBs observed by the HETE-2 satellite \cite{doty} as a function of the redshift and look for a dependance of $\Delta t$ as a function of $z$:
\begin{equation}
< \Delta t >\ = \mathrm{a}\,K_l(z) + \mathrm{b}\,(1+z),
\label{form:dtkl}
\end{equation}
where 
$$K_l(z) = \int_{0}^{z} \frac{(1+z)\,dz}{h(z)}$$
and $\mathrm{a}$ and $\mathrm{b}$ parameters stand for extrinsic (Quantum Gravity) and intrinsic effects, respectively.

As in \cite{ellis1, ellis2}, we use wavelet analysis \cite{mallat} to study the light curves. Unlike Fourier transform, this tool is well adapted to the study of non-stationary signals, \textit{i.e.} signals for which the frequency changes in time.

In this paper, we will only discuss maxima of the light curves. A study of minima is presented in \cite{jb} along with a more detailed description of the ana\-lysis. In the next section, we describe the method we use to study the light curves and to measure the time lags. Then, we give our results in terms of a limit on the quantum gravity energy scale $\mathrm{E}_\mathrm{QG}$. The intrinsic effects are related to the source itself. They are discussed in the last section.

\begin{figure}[t!]
	 \centering
   \includegraphics[width=0.45\textwidth]{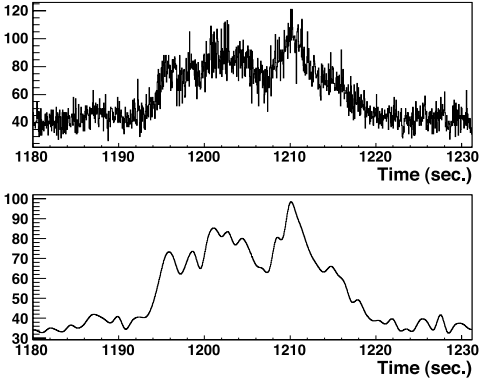} 
   \caption{Light curve of \texttt{GRB 041006} for the energy band 8-30~keV before (top) and after (bottom) denoising.}
   \label{fig:denoise}
\end{figure}

\begin{figure}[t!]
	 \centering
   \includegraphics[width=0.45\textwidth]{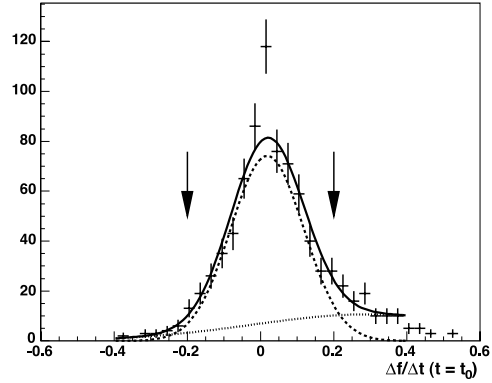} 
   \caption{Distribution of the derivatives of the light curve for maxima located at $t_0$. The contribution of fake extrema is negligible for $|\Delta f/\Delta t| \leq 0.2$.}
   \label{fig:der}
\end{figure}

\section{Method}

The analysis of the 15 GRBs with measured redshift follows the steps described below:

\begin{itemize}
\item
determination of the time interval to be studied between start and end of burst. It is defined by a cut above the background measured outside of the burst region,
\item
choice of the two energy bands for the time lag calculations, later called energy scenario, by assigning the individual photons to each energy band. 14 different scenarios were considered, corresponding to 14 different values of the energy gap between low and high energy bands,
\item
de-noising of the light curves by a Discrete Wavelet Transform (DWT, see Fig.~\ref{fig:denoise}) and pre-selection of data in the studied time interval for each GRB and each energy band,
\item
search for the rapid variations (spikes) in the light curves for all energy bands using a Continuous Wavelet Transform (CWT). The result of this step is a list of maxima candidates, along with a coefficient characterizing their regularity (Lipschitz coefficient $\alpha$ and its error $\delta\alpha$),
\item
association in pairs of the maxima, which fulfill the conditions derived from studies of the Lipschitz coefficient.
\end{itemize}

\begin{figure}[t!]
	 \centering
   \includegraphics[width=0.45\textwidth]{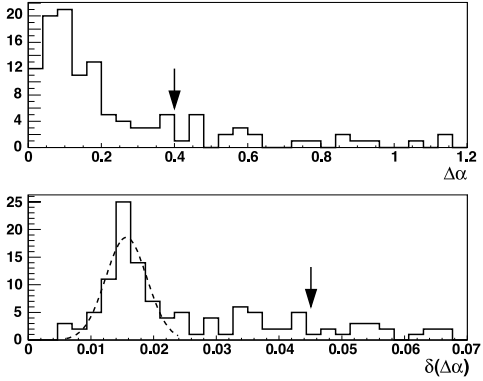} 
   \caption{Distributions of $\Delta\alpha$ (top) and $\delta(\Delta\alpha)$ (bottom). Arrows give the positions of the cuts.}
   \label{fig:sigalpha}
\end{figure}

Different cuts and selections are performed at different steps in the analysis. A cut on the derivative of the light curve at the position of each local extremum is done to ensure that it is not fake (Fig.~\ref{fig:der}). Then, for each pair (indexed by 1 and 2 for low- and high-energy band, respectively), the following quantities are computed:
$$
\left\lbrace
\begin{array}{l}
\Delta t = t_2 - t_1\\
\Delta\alpha = | \alpha_2 - \alpha_1 |\\
\delta(\Delta\alpha) = \sqrt{\delta\alpha_2^2 + \delta\alpha_1^2}. \\
\end{array}
\right.
$$
and the following selections are applied:
$$
\left\lbrace
\begin{array}{l}
\Delta\alpha < 0.4 \\
\delta(\Delta\alpha) < 0.045. \\
\end{array}
\right.
$$
These cuts are based on the distributions shown in Fig.~\ref{fig:sigalpha} and are valid for all the energy gap scenarios we consider.

As a result, a set of associated pairs is produced for each GRB and each energy scenario. The average time lag of each GRB, $<\Delta t>$, is then calculated and used later in the study of the Quantum Gravity model described in the previous section. The evolution of the time lags as a function of the redshift $z$ allows us to constrain the minimal value of the Quantum Gravity scale $\mathrm{E}_\mathrm{QG}$.

\section{Results}

\begin{figure}[t!]
	 \centering
   \includegraphics[width=0.47\textwidth]{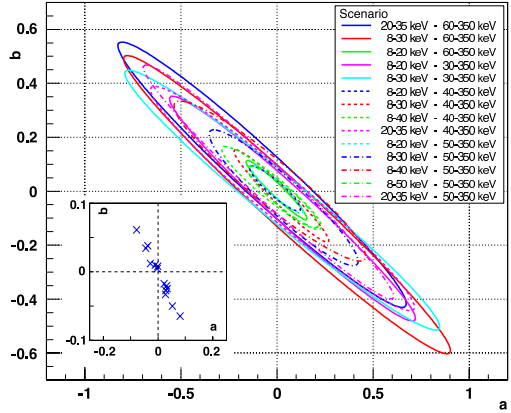} 
   \caption{95\% contours for $\mathrm{a}$ and $\mathrm{b}$ for the two-parameter fit for the 14 scenarios for maxima. The box at the bottom left shows the position of contour centers.}
   \label{fig:cont}
\end{figure}

The two parameters $\mathrm{a}$ and $\mathrm{b}$ of Eq.~\ref{form:dtkl} were fitted for the 14 energy scenarios. Both parameters were found to be strongly correlated, as shown by Fig.~\ref{fig:cont}. However, results of the fits do not show any significant deviation of $\mathrm{a}$ from 0 at $\pm3\sigma$, so in the following, we derive the 95\% CL limit on the Quantum Gravity scale $\mathrm{E}_\mathrm{QG}$.

For this, we define a likelihood function $L$ by
$$L = \exp\left(-\frac{\chi^2(M)}{2}\right),$$
where $M$ is the energy scale and where $\chi^2(M) =$
\begin{equation}
\label{form:chi2}
\sum_{\mathrm{all\ GRBs}} \frac{\left(\Delta t_i - \tilde{b}(1+z_i) - a_i(M)\,{K_l}_i\right)^2}{\sigma_i^2 + \sigma_{\tilde{b}}^2},
\end{equation}
where the index $i$ corresponds to each GRB. The dependence of $a_i$ on $M$ as predicted by the considered model of Quantum Gravity is given by: 
$$a_i(M) = \frac{1}{H_0}\,\frac{\Delta <E>_i}{M},$$
where $\Delta <E>_i$ is evaluated for each GRB using the relation
$$\Delta<E>\ =\ <E>_2 - <E>_1.$$

Assuming a universality of the intrinsic time-lags, the average value $\tilde{b}$ (and its error $\sigma_{\tilde{b}}$)  of Eq.~\ref{form:chi2} was obtained as the weighted mean of the values $b_k$ (and their errors $\sigma_k$) from the two-parameter linear fit:
$$\tilde{b} = \frac{\sum_k w_k\,b_k}{\sum_k w_k}\ \ \mathrm{and}\ \ \sigma_{\tilde{b}} = \frac{1}{\sqrt{\sum_k w_k}},$$
where the index $k$ corresponds to each scenario and $w_k = 1/\sigma_k^2$.
For maxima, we obtain $\tilde{b} = 0.0023\pm0.0027$.

\begin{figure}[t!]
	 \centering
   \includegraphics[width=0.47\textwidth]{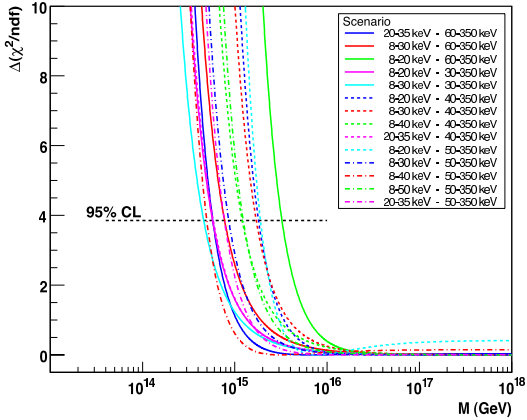} 
   \caption{Evolution of $\chi^2$ as a function of $M$.}
   \label{fig:chi2}
\end{figure}

Figure~\ref{fig:chi2} presents the evolution of $\chi^2(M)/\mathrm{ndf}$ around its minimum for maxima of the light curves. All scenarios fulfill the condition $\chi_{min}^2/\mathrm{ndf} \le 2$.
No significant preference of any value of $M$ is observed for most of the scenarios.

The 95\%~CL lower limit on the Quantum Gravity scale was set by requiring $\chi^2/\mathrm{ndf}$ to vary by 3.84 from its minimum.
The values of limits on $\mathrm{E}_\mathrm{QG}$ were obtained for the 14 energy scenarios. In good agreement with the slope parameter $\mathrm{a}$, all values for maxima are within the $10^{14}$-$10^{16}$~GeV range. The scenario which provides the strongest constraint on the QG scale (\mbox{8-20~keV}, \mbox{60-350~keV}) provides us a lower limit of \mbox{$\mathrm{E}_\mathrm{QG} > 3.2\times10^{15}$~GeV}.

\section{Discussion}

GRB light-curves are not ``perfect'' signals, since they exhibit intrinsic time lags between high and low energies. The peaks of the emission are shorter and arrive earlier at higher energies \cite[and references therein]{fenimore,norris6}. These intrinsic lags have a sign opposite to the sign expected from Lorentz violation. The effect of Lorentz violation must therefore be searched with a statistical study analyzing the average dependence of the lags with $z$, and not in samples limited to one GRB.

A strong anticorrelation of spectral lags with luminosity has been found in \cite{norris}. At low redshifts, we detect bright and faint GRBs which present a broad distribution of instrinsic lags whereas at high redshifts, we detect only bright GRBs with small intrinsic lags. This effect could mimick the effect of Lorentz violation due to the non uniform distribution in luminosity of the GRBs in our sample. Here, with limited statistics, we made a test by performing our analysis on a restricted GRB sample, almost homogeneous in luminosity. The obtained results show lower sensitivity (lower values of the limits on Quantum Gravity), because of a decreased statistical power of the restricted sample and a smaller lever arm in the redshift values.

\section{Conclusion}

Light curves of 15 GRBs with known redshifts observed by the satellite HETE-2 have been studied using wavelet analysis in order to look for a Quantum Gravity effect on light propagation. No effect is detected above $\pm3\sigma$ and a lower limit on $\mathrm{E}_\mathrm{QG}$ is set to $3.2\times10^{15}$~GeV. This limit can be considered as competitive considering the energy gap of $\sim$130~keV provided by HETE-2.

\end{document}